%% file: main.tex
\documentclass[USenglish,oneside,twocolumn]{article}

\usepackage[utf8]{inputenc}
\usepackage[big]{dgruyter_NEW}
\usepackage{adjustbox}
\usepackage{pifont}
\usepackage{rotating}
\newcommand{\cmark}{\ding{51}}%
\newcommand{\xmark}{}%

\newcommand{\para}[1]{\noindent\textbf{#1}} 
\newcommand{\define}[2]{\vspace{0.75mm}\noindent\textbf{#1}: \textit{#2}\vspace{0.75mm}} 

\newcolumntype{R}[2]{%
    >{\adjustbox{angle=#1,lap=\width-(#2)}\bgroup}%
    l%
    <{\egroup}%
}
\newcommand*\rot{\multicolumn{1}{R{25}{1em}}}

\newcommand{\categrow}[1]{
    \cline{1-13}
    \multicolumn{13}{l}{#1} \\
    \cline{1-13}
}

\DOI{foobar}
\cclogo{\includegraphics{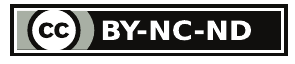}}

\begin{document}

\title{\huge Systematizing Decentralization and Privacy: Lessons from 15 Years of Research and Deployments}

\runningtitle{Systematizing Decentralization and Privacy}

\author[1]{Carmela Troncoso}
\author[2]{Marios Isaakidis}
\author[3]{George Danezis} 
\author[4]{Harry Halpin}

\affil[1]{IMDEA Software Institute, E-mail: carmela.troncoso@imdea.org}
\affil[2]{University College London, E-mail: m.isaakidis@cs.ucl.ac.uk}
\affil[3]{University College London, E-mail: g.danezis@ucl.ac.uk}
\affil[4]{INRIA, E-mail: harry.halpin@inria.fr}

\journalname{Proceedings on Privacy Enhancing Technologies}
\DOI{10.1515/popets-2017-0052}
  \startpage{307}
  \received{2017-02-28}
  \revised{2017-06-01}
  \accepted{2017-06-02}

  \journalyear{2017}
  \journalvolume{}
  \journalissue{4}

\begin{abstract}{Decentralized systems are a subset of distributed systems where multiple authorities control different components and no authority is fully trusted by all. This implies that any component in a decentralized system is potentially adversarial. We revise fifteen years of research on decentralization and privacy, and provide an overview of key systems, as well as key insights for designers of future systems. We show that decentralized designs can enhance privacy, integrity, and availability but also require careful trade-offs in terms of system complexity, properties provided, and degree of decentralization. These trade-offs need to be understood and navigated by designers. We argue that a combination of insights from cryptography, distributed systems, and mechanism design, aligned with the development of adequate incentives, are necessary to build scalable and successful privacy-preserving decentralized systems.}
\end{abstract}

\maketitle

\section{Introduction: the Long Road from 2001 to 2016}

\input{s01intro}

\section{Epistemology}
\input{s02background}

\section{Decentralization and Privacy}
\input{s030Questions}
\section{Future Research Lines}
\input{s04directions}

\section{Conclusions: Towards Full Decentralization}

\input{s05conclusions}

\vspace{2mm}
\noindent \textbf{Acknowledgements.} The authors would like to thank the reviewers for insightful comments that helped improving the paper, in particular Prateek Mittal for acting as shepherd. This work is supported by the EU H2020 project NEXTLEAP (GA 688722).

\bibliographystyle{abbrv}
\bibliography{SoKBib}

\end{document}

%% file: s01intro.tex
The successful adoption of decentralized systems such as BitTorrent~\cite{bittorrent}, Tor~\cite{DingledineMS04}, and Bitcoin~\cite{nakamoto2008bitcoin}, and the revelations of mass surveillance against centralized cloud services~\cite{greenwald2014no}, has contributed to the wide belief that decentralized architectures are beneficial to privacy.
Yet, there does not exist a foundational treatment or even an established common definition of decentralization. 
In this paper we aim at defining decentralization and systematizing the ways in which a system can be decentralized, and, by presenting the key design decisions in decentralized systems, bring forth past lessons that can inform a new generation of decentralized privacy-enhancing technologies.

This is not the first time there has been a surge of interest in decentralization. As Cory Doctorow noted at the 2016 Decentralized Web Summit: ``It's like being back at the O'Reilly P2P conference in 1999,''
which signaled a peak of interest around decentralized architectures at the turn of the millennium~\cite{oram2001peer}. The `hype' around decentralization was followed in the early 2000s by research and deployment activity around decentralized systems.

To some extent, decentralization was originally a response to the threat of censorship. Perhaps the first rallying cry for decentralization was the Eternity Service~\cite{Anderson96}. Anderson created this system in response to the success of the Church of Scientology at closing down the anon.penet.fi remailer~\cite{Helmers97} ``as a means of putting electronic documents beyond the censor's grasp.''
This motivation of censorship resistance is clear in more modern systems: Tor using a decentralized network of anonymous relays and a DHT-based hidden services naming infrastructure; Bitcoin emerging as a censorship-resistant way to transfer funds to organizations like Wikileaks after the centralized e-Gold \cite{egold} online currency had been shut down by the Department of Justice; or BitTorrent succeeding as a peer-to-peer (P2P) file sharing service using Mainline DHT~\cite{WangK13} rather than having a central indexing service like Napster that could be subject to requests to keep track of file copying~\cite{napstercase}. In each of these cases, decentralization arose as a response to the shutdown of a centralized authority, aiming to remove that single natural point of failure.

Despite the millennial fervour for decentralization, the 2000s witnessed the rise of massively distributed, \textit{but not decentralized}, data centers and systems as the dominant technical paradigm embodied by the Cloud computing capabilities offered by Google, Facebook, Microsoft, and others. Eventually, users were diverted away from software running locally on their machines, which essentially is a form of decentralization, towards cloud applications that enabled an unprecedented aggregation of user data by the providers. Snowden's revelations in 2013 on mass surveillance programs leveraging the centralized nature of these services gave credence to long-standing privacy concerns brought about by the rise and popularity of centralized services. 

The desire to preserve privacy, liberty, and the autonomous control of infrastructure and services have led to a call to ``re-decentralize'' the Internet~\cite{redecentralize,ybti}. As a result, in the 2010s we are observing an upsurge of alternatives to centralized infrastructures and services, although most alternatives to Cloud-based applications are still under development.

It is important for system designers to neither be nostalgic about past systems nor fatalistic about future ones. Today's networking and computing environments are vastly different from those in 2000: Smart-phones have placed a powerful computer in people's pockets; users are usually connected to the Internet over fast connections without time or bandwidth caps; clients, such as web browsers, are now mature end-used platforms with P2P communications enabled and cryptographic capabilities; and mobile code, in the form of Javascript, is ubiquitous. 

Even though the design space for modern decentralized systems is less restricted than in the past, fundamental challenges remain. Our key objective is to support future work on decentralized privacy systems by systematizing the past 15 years of research, between O'Reilly's publication of ``Peer-to-Peer: Harnessing the Power of Disruptive Technologies''~\cite{oram2001peer} in 2001, and 2016. We aim at highlighting key findings in classic designs, and also the important problems faced by designers of past systems, so as to inform the choices made by engineers pursuing decentralization today.

%% file: s02background.tex
\noindent
\para{Scope.}
There is a wide use of the term `decentralized'. In this paper, we restrict ourselves to discussing systems that support privacy properties using decentralized architectures. We draw a distinction between \textit{decentralized} and \textit{distributed} architectures, as follows:

\define{Distributed system}{A system with multiple components that have their behavior co-ordinated via message passing. These components are usually spatially separated and communicate using a network, and may be managed by a single root of trust or authority.} 
Distribution is beneficial to support robustness against single component failure, scalability beyond what a single component could handle, high-availability and low-latency under distributed loads, and ecological diversity to prevent systemic failures. Developments led by Google, ranging from BigTable~\cite{chang2008bigtable} to MapReduce~\cite{dean2008mapreduce} are good examples of distributed systems.

\define{Decentralized system}{A distributed system in which multiple authorities control different components and no single authority is fully trusted by all others.}

\begin{figure}
\centering
 \includegraphics[width=\linewidth]{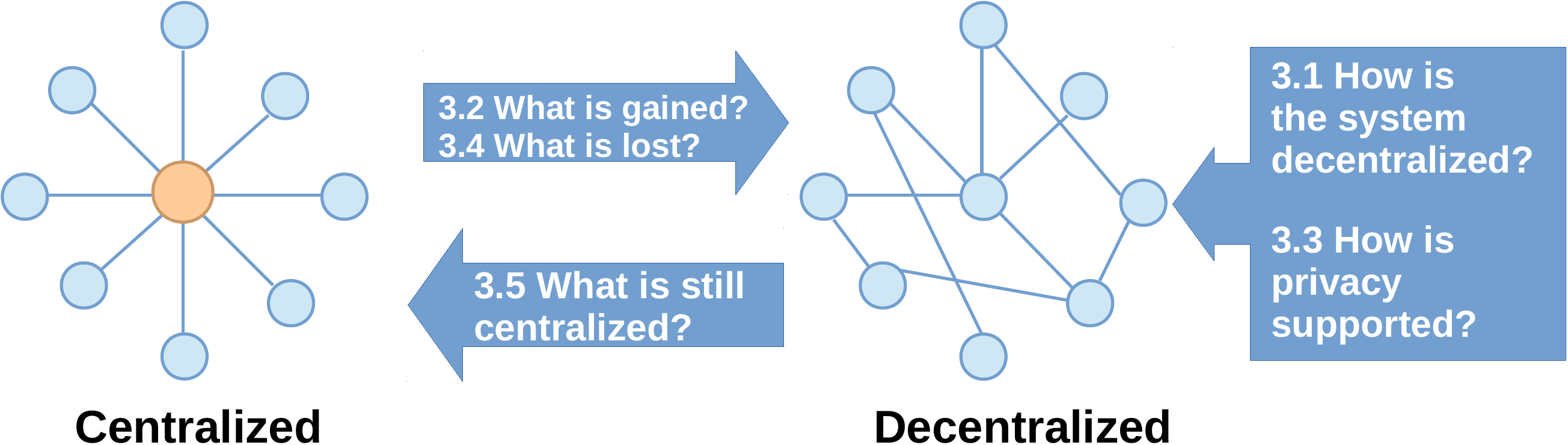}
\caption{From centralized to decentralized systems}
\label{fig:decen}
\end{figure}

Following Baran~\cite{baran1964distributed}, systems are conceived of as networks of interconnected components, where all the components of a system form a graph, where the nodes of the graph are the components and the edges the connections between them (see Fig.~\ref{fig:decen}).
Due to this analogy with graphs, the terms ``decentralized network'' and ``decentralized system'' tend to be used interchangeably. However, decentralized systems are not just network topologies, but systems that exist to fulfill some function or set of functions, otherwise called `operations.' These operations are accomplished by passing messages between a sender and a receiver node, with other nodes serving as proxies to relay the message~\cite{lamport1978time} (right graph in Fig.~\ref{fig:decen}). On the contrary, in centralized systems messages and operations are orchestrated by a central trusted authority (depicted as an orange circle in the left graph in Fig.~\ref{fig:decen}). 

Centralized systems may be distributed, typically for efficiency or scaling, but not for privacy, and so the underlying components are fundamentally trusted. Only external entities are considered adversarial. 
Widely deployed systems such as Bitcoin, BitTorrent, and Tor are on the other hand decentralized. Contrary to generic distributed systems, in participating parties may choose their relationships of trust autonomously, including the case where there one may not trust any other components. This has profound implications in terms of security and privacy: no single entity that can act as a trusted computing base (TCB)~\cite{rushby1981design} to enforce a global security or privacy policy. Any internal component of the system may be adversarial, in addition to external parties, requiring defences in depth.

In terms of security and privacy we adopt the following broad definitions, that we make more detailed at the corresponding section when the context requires clarification or preciseness.

\define{Security}{We consider the security aspects of a system to be those that encompass traditional information security properties. This include of course confidentiality, integrity, and authentication; but also less traditional ones such as availability, accountability, authorization, non-repudiation or non-equivocation.}

\define{Privacy}{We consider the privacy aspects of a system to be those related to the protection of users' related data (identities, actions, etc.). This protection is usually formalized in terms of privacy properties (anonymity, pseudonymity, unlinkability, unobservability) for which we follow the definitions by Pfitzmann and Hansen~\cite{PfitzmannH05}. These definitions are extended in the privacy-oriented discussion in Section~\ref{sec:privacy}}.

\vspace{2mm}
\para{Methods \& Model.}
To systematize knowledge in decentralized privacy-preserving systems we performed a systematic literature review of all papers published in the top 4 computer security conferences (IEEE S\&P, ACM CCS, Usenix Security, NDSS) as well as the specialized conferences (PETS, WPES and IEEE P2P) that are proposing or analyzing decentralized systems with privacy properties, from the years 2000 to 2016.

Our first analysis resulted in 165 papers (28 from IEEE S\&P, 56 from ACM CCS, 18 from Usenix Security, 11 from NDSS, 11 from PETS, 10 from WPES, and 31 from IEEE P2P). Finally the paper contains only 90 references from these venues (13 from IEEE S\&P, 32 from ACM CCS, 10 from Usenix Security, 11 from NDSS, 9 from PETS, 6 from WPES, and 9 from IEEE P2P), 19 are well-known deployed systems that do not have an associated peer-reviewed publication, and the rest come from an additional pool of 30 conferences and workshops (among them FOCI, WEIS, NSDI, SIGCOMM, SIGSAC, or CRYPTO). The selection was done on the basis of highlighting design decisions that reflect a key lesson worth of future reference.

Due to the vast amount of identified designs, by necessity we do not describe each system in detail, but instead show how each system exemplifies a property or design choice. We do, though, expand upon Tor, BitTorrent, and Bitcoin as they are are heavily deployed and have substantial academic analysis. As illustrated in Figure~\ref{fig:decen}, we study the pool of selected designs with the intention to determine:
\begin{enumerate}
    \item How is the system decentralized? (Section~\ref{sec:how})
    \item What advantages do we get from decentralizing? (Section~\ref{sec:gain})
    \item How does decentralization support privacy? (Section~\ref{sec:privacy})
    \item What are the disadvantages of decentralizing? (Section~\ref{sec:lose})
    \item What implicit centralized assumptions remain? (Section~\ref{sec:assumptions})
    \item What can we learn from existing designs? (Section~\ref{sec:table})
\end{enumerate}

\vspace{2mm}
\para{Insights}. 
\begin{itemize}
    \item \textit{The key difference between distributed systems and decentralized systems is one of authority and trust between components. Differences in architecture and use of security and privacy controls stem from it.}
    \item \textit{Decentralized systems embody a complex set of relationships of trust between parties managing different aspects of the system. Untrusted insiders are common, and security controls must be deployed taking into account adversaries within the system.}
    \item \textit{In distributed, but not decentralized, systems the existence of a single authority that provisions and manages all components that are trusted enables the use of simple security, many times based on dedicated trusted components that act as roots of trust.}
    \item \textit{In decentralized systems no single authority can provision a root of trust or trusted computing base, making security mechanisms reliant on those (such as central access control or traditional public key infrastructures) inapplicable. }
\end{itemize}

%% file: s030Questions.tex
This section runs over the key questions we pose in the previous sections with regards to the current state of affairs in decentralized systems. Table~\ref{table:systemsTable} (page \pageref{table:systemsTable}) provides a summary of the different design decisions and the properties achieved as a result.

\subsection{How Is Decentralization Achieved?}
\input{s031type}

\subsection{The Advantages of Decentralization}
\input{s032gain}

\subsection{How Does Decentralization Support Privacy?}
\input{s033privacy}

\subsection{The Disadvantages of Decentralization}
\input{s034lose}

\subsection{What Is Still Centralized in Decentralized Designs?}
\input{s035assumptions}

\subsection{Systematization of Existing Designs}
\label{sec:table}

\input{s036table}
\AtBeginShipout{
\ifnum\thepage=12
\pdfpageattr{/Rotate 90}
\fi
}

Table \ref{table:systemsTable} presents a systematic analysis of decentralized designs, clustered based on their principal goal. The columns infrastructure, network topology, authority relations, privacy properties, follow closely the definitions of the previous subsections. We applied some level of simplification to complex systems with multiple components or multiple use-cases. The systematization focuses on parts of the system relevant for its main use-case as used in prototype or deployment. 

\vspace{2mm}
\para{Insights}.
\begin{itemize}
   \item \textit{Many systems that provide good coverage of privacy properties and decentralization (usually via DHTs) have not been widely deployed}
   \item \textit{Widely deployed systems either are user-independent federated systems or user-based DHT-based systems, both without advanced privacy properties.}
   \item \textit{Hybrid and stratified systems such as Tor provide provide advanced privacy properties at the cost of centralized assumptions.} 
   \item \textit{The space of ad-hoc, mesh, and covert designs is under-explored.} 
\end{itemize}

%% file: s031type.tex
\label{sec:how}

We review key architectural decisions: how to orchestrate the infrastructure of the network, how to route messages, and how to distribute trust between nodes.

\subsubsection{Infrastructure}
A first key choice concerns the distribution of tasks needed for maintaining a service within the system. The provisioning of infrastructure impacts the design in terms of trust and message routing. 

\vspace{2mm}
\para{User-based Infrastructure}. Some decentralized system consist solely of nodes that are users and there is no additional infrastructure. They rely solely on users to collectively contribute resources (bandwidth, storage) in order to provide a service. The advantage of this design is that by nature it does not require a third-party centralized authority. This user-based design can support services such as hosting of encrypted data, e.g. in Freenet~\cite{ClarkeSWH00} and Cachet~\cite{NilizadehJMBK12}. 
A disadvantage is that user-based infrastructure may lead to poor performance due to evolving into sparsely connected topologies, and to ``churn'' caused by peers constantly joining and leaving the network.

\vspace{2mm}
\para{User-independent Infrastructure}. Here, the functions of the decentralized system are realized by nodes that are not users. A set of third-parties that are not necessarily trusted may provide all or part of the functionality to users. This design pattern underlies classic open federated protocols such as SMTP~\cite{smtp} and XMPP~\cite{xmpp} based on a client-server model. The advantages of user-independent infrastructure include increased availability of the service, a reduced attack surface, and immunity to user churn. Servers do not necessarily threaten user privacy. The Eternity Service~\cite{Anderson96}, as realized in systems like Tahoe-LAFS~\cite{SelimiF14}, combined encryption with the use of several servers controlled by different non-collaborating authorities for the private storage and replication of files. Other examples of systems that rely on user-independent infrastructure include DP5~\cite{BorisovDG15} and Riposte~\cite{Corrigan-GibbsB15} in terms of Private Information Retrieval~\cite{ChorKGS98} or anonymous communication systems like mix networks~\cite{Chaum81} or DC-nets~\cite{Chaum88}. 

\vspace{2mm}
\para{Hybrid Systems}. Functions may be shared between users and nodes run by third-parties. An example is Tor, where relays are mainly run by volunteers but Directory Authorities are operated by a closed `known' group of servers. 
In terms of privacy and security, new elements such as distributed ledgers decentralize traditionally centralized cryptographic protocols in these hybrid systems. For example, computations can be locally and securely recorded to the blockchain with the support of multi-party computation protocols~\cite{ZyskindNP15}, even without a trusted third party~\cite{AndrychowiczDMM14,ZyskindNP15}, or using a small number of stable entities to ensure reliability and low-latency, as in the Sharemind MPC system~\cite{bogdanov2008sharemind}.

\subsubsection{Network Topology}
\label{topology}
When considering a decentralized system, there are two distinct topologies. The first, \textit{network topology} describes the connections between nodes used to route traffic; and the second, \textit{authority topology} describes the power relations between the nodes. Thus, the network routing structure does not necessarily have to mirror how authority is decentralized in a system, although it often does. That can greatly affect the security and privacy properties of the system~\cite{DiazMT10}.
It must be noted that components of traditional network routing is done in a hierarchical manner, including spanning tree protocols such as in BGP~\cite{rekhter2005border} in the current Internet as well as `next generation' designs like SCION~\cite{zhang2011scion}.

\vspace{2mm}
\para{Mesh}. Mesh topologies are unstructured. Nodes can route messages to every other node they are connected with. One advantage is that mesh networks function in settings with no stable connections to other nodes to guarantee service in the presence of massive churn and changing connectivity, such as in mobile ad-hoc networking and file sharing in early versions of Gnutella~\cite{Gnutella}. A particularly popular communication means in mesh topologies~\cite{nakamoto2008bitcoin} are \textit{gossip protocols}. In gossiping, as opposed to flooding, a random subset of the nodes in the network are chosen to receive the messages. These nodes then continue to broadcast the message via another independently selected random subset of the network to relay messages. The reliability of message delivery under load is questionable and information propagation experiences delays. Historically mesh networking does not preserve user privacy of their users, but recent secure messaging systems such as Briar~\cite{briar} use this topology to remain functional during Internet blackouts.

\vspace{2mm}
\para{Distributed Hash Tables} (DHT). DHTs are network topologies where each node maintains a small routing table of its neighbours, and messages are passed greedily to known nodes that are `closer' to the intended recipient. Although efficient and decentralized, DHTs do not by themselves provide strong security, privacy and anonymity properties. While decentralized, DHTs are not secure and privacy-preserving by default: Tran et al.~\cite{TranHK09} show that low latency anonymity systems based on DHTs such as Salsa~\cite{NambiarW06} are vulnerable to having large amounts of traffic captured by adversaries controlling a fraction of the relays. DHT nodes may, however, be grouped into byzantine quorums to defeat adversaries that control a minority of nodes~\cite{young2010practical}.

\vspace{2mm}
\para{Super-nodes}. Super-nodes are nodes that are endowed with more, and contribute more, resources to the system. This may be in terms of computation power, storage, or network connectivity, stability and up time. In terms of routing, such super-nodes may be used to mediate operations requiring higher network throughput. They can be arranged in structured topologies, designed to leverage them; or they may emerge naturally in unstructured topologies, as a result of some nodes committing more resources. Most P2P systems such as BitTorrent eventually rely on super-nodes~\cite{DeckerEW13}. These super-nodes have serious implications on availability and integrity, as they may become targets for attack, and privacy, as they mediate, and are in a privileged position to observe, a larger fraction of activities.

\vspace{2mm}
\para{Stratified}. Some of the more complex decentralized systems use a stratified design where nodes have specialized roles in terms of routing, or other functions. A paradigmatic example is the Tor network. Tor users autonomously form circuits from an open-ended set of Tor relays, in layers of entry guards, middle nodes and exit nodes. A high-integrity global list of these relays is maintained through consensus by a closed group of specialized Directory Authorities. Simultaneously, Tor hidden services are resolved through a Hidden Service Directory maintained by a simple DHT topology. We note that, on some level, Tor has also evolved to use super-nodes on its topology and the distribution of traffic sent through Tor relays is far from uniform~\cite{JohnsonWJSS13}. Cascades, are a particular case of Stratified topologies in anonymous communications, in which paths are pre-defined. The advantages and disadvantages of such choice as opposed to free routes has been discussed in~\cite{DiazDGPS04}.

\subsubsection{Authority} \label{sssec:authority}
We now consider the relation among nodes in terms of authority and describe mechanisms to mitigate the potentially effects of power disparity that could potentially harm the security and privacy of users.  

\vspace{2mm}
\para{Ad-hoc: Nodes Interact Directly}. In ad-hoc there is no relationship of authority among nodes. Nodes directly interact with each other without the participation of other nodes, and they do so for the benefit of the involved parties only. In terms of routing, ad-hoc requires a mesh topology where nodes do not carry traffic for other nodes. However, note that mesh topologies do not always have a ad-hoc (lack of) authority relations, such as routing based on gossip. An example of this type of system would be point-to-point communication in Briar \cite{briar}. 
For purposes of privacy, direct interaction bypasses possibly compromised nodes, but not network adversaries. As for confidentiality, communications can be encrypted between the two nodes, and can be extended to group communication using group key agreement protocols~\cite{SchmidtSCB14}.

\vspace{2mm}
\para{P2P: Nodes Assist Other Nodes}. P2P designs have no central authority. Unlike ad-hoc interaction, nodes provide services and resources to other nodes, such as routing messages or storing blocks of data. Nodes have equal authority and so each node may equally compel any other node, although services and resources are usually provided according to their capacity. In other words, P2P systems self-organize and all nodes are responsible for carrying out operations for all other nodes, rather than having any pre-configured special position of authority. Since nodes are not motivated by authority to help each other, mechanisms should instead be in place to provide `incentives' for collaborative behaviour.

There are clear advantages for the security and privacy properties in P2P systems. Information about peers is not centralized and interaction typically remains local to a few nodes, so it is difficult for an adversary to obtain a global view of the system. Yet, relying on peers for functionality poses an additional threat to privacy, since requests may be served by adversarial nodes. These nodes can passively collect information on other nodes or they may actively disrupt the integrity of operations by forging messages or replay attacks that are hard to detect. Furthermore, since P2P systems are usually open, without any admissions control, adversaries may purposely inject a large number of Sybil nodes, to increase their chances of a successful attack~\cite{Douceur02}. P2P systems are not a silver bullet for decentralization: there is no clear and definite solution to Sybil attacks in P2P networks, although such an attack can be mitigated using reputation~\cite{DamianiVPSV02} or trust~\cite{JohnsonSDM11}.

\vspace{2mm}
\para{Social-based: Nodes Assist Friends}. These designs take advantage of pre-existing decentralized relationships, such as ``friendship''. 
In terms of applicability of security mechanisms this approach maintains most advantages of a P2P system. It is less vulnerable to Sybil attacks as adversarial nodes can be excluded from participating in the network or may be easier to detect~\cite{DanezisM09}, as it is harder to infiltrate a social network than a network. The downside is that, without cover traffic, a global passive adversary can discover the underlying social graph by monitoring network communications and violate privacy properties such as unobservability and unlinkability. This in turn may lead to user deanonymization~\cite{NarayananS09}, and techniques such as perturbation of the underlying graph may not be robust enough to prevent this~\cite{mittal2012preserving}.

A number of systems implement social-based communication to resist Sybil attacks. For instance Drac~\cite{DanezisDTL10} and Pisces~\cite{MittalWB13} use social-networks to support routing of messages. X-Vine~\cite{MittalCB12} is a mechanism that, applied to distributed hash tables, helps resisting denial of service via Sybil attacks at the cost of higher latency. Tribler~\cite{PouwelseGWBYIERSS06} uses social-based trust relations to improve performance that exploits similarity to improve performance, content discovery, and downloading in file sharing; or Nasir et al.'s socially-aware DHT~\cite{NasirGK15}, which reduce latency and improve the reliability of the communication.

\vspace{2mm}
\para{Federated: Providers Assist Users}. In federated designs, users are associated to \textit{provider} nodes, which they trust and that act as authorities. Each provider is responsible only for its own users but collaborates with other providers in order to provide a service. No single provider has authority over other providers, and thus there is a ``federation'' of providers. Federated authorities typically use user-independent infrastructure and act as a super-node in terms of routing. This combination of design choices leads typically to high availability as long as the provider is accessible and not compromised, but the provider is a central point of attack to violate security properties and the provider itself can violate the privacy of nodes.
The primary weakness of federated systems is the assumption that federated service providers largely act honestly. Some techniques can relax strong trust assumptions in the provider. End-to-end encryption can maintain confidentiality~\cite{sparrow2016leap} using providers. Computation can be obscured using secret sharing~\cite{RogawayB07} or differential privacy-based solutions~\cite{AlhadidiMFD12}.

\vspace{2mm}
\para{Accountability: Transparency Assists Users}. \label{sssec:type-accountability} Transparency can be used to make an authority accountable in order to establish trust. It promotes integrity of operations by monitoring the correct behavior of nodes, e.g. a transparent log of a provider's operations in a federated system audited by users or other providers acting in lieu of their associated users. The nature of this auditor's authority is very different from the aforementioned previous types of authority relations and critically relies on the non-collusion of the auditor and the audited authority, e.g., Bitcoin consensus over its blockchain using proof-of-work.
Other alternatives, such as Certificate Transparency~\cite{laurie2014certificate}, rely on a set of services and auditors to keep track of X.509 certificates and quickly detect potentially rogue or hacked certificate authorities. Similarly, electronic election protocols~\cite{gritzalis2012secure} achieve robustness through proofs of correct shuffling of votes, e.g., Helios~\cite{Adida08}. Yet na{\"i}ve designs of audit logs may violate the privacy of decentralized nodes by learning too much information. 

While decentralized accountability can have clear advantages regarding integrity, there are difficulties in maintaining privacy in any distributed log. This disadvantage can nevertheless be reduced as shown by Zerocash~\cite{Ben-SassonCG0MTV14}, which uses zero-knowledge proofs in order to maintain unlinkability in auditing relationships; or CONIKS~\cite{MelaraBBFF15}, that shows that auditing the consistency of a name-key binding through time enables verification of user public keys by the end users collectively and by other providers, while concealing the identities and the number of users at each provider using Verified Random Functions.

\vspace{2mm}
\para{Insights}. 
\begin{itemize}
    \item \textit{Decentralization encompasses a large space of designs from decentralized ad-hoc mesh to federated super-node networks, not just peer-to-peer. These offer a variety of privacy and systems (e.g., availability, or reliability) properties. Developer instincts may often be incorrect in terms of their trade off.}
    \item \textit{Despite being separate parts of the design, the network topology in decentralized systems often mirrors the authorities' trust relationships. However, a strict mapping between authority, infrastructure and networking topology is not necessary, and may come at the cost of harming privacy or availability.}
    \item \textit{Centralization in terms of federated and super-nodes leads to better availability and system performance. However, it introduces single points of failure that impact availability and privacy. P2P models are by design more resilient to unstable routing and compromises, but entail higher engineering complexity.}
    \item \textit{All networking topologies suffer under node churn, and pure P2P topologies must effectively address this effectively to be applicable at all.}
    \item \textit{Decentralization does not imply the absence of any infrastructure. However, the infrastructure itself needs to be decentralized by being provided by a plurality of authorities. Such infrastructure may enhance performance by offering super-nodes or dedicated high-availability operations.}
    \item \textit{De-facto super nodes may emerge naturally in decentralized designs, as a result of different node capabilities, and efficiency in centralizing certain operations. If this occurs outside the context of careful design, those super nodes become a single point of failure, and may lead to de facto re-centralization.}
    \item \textit{Lack of relationships of authority imply that nodes must be willing to provide services to each other on a different basis. Designers of decentralized systems must carefully engineer such incentives, to ensure that natural (non adversarial) selfishness does not lead to dysfunction. Monetary incentives, reputation, and reciprocity can be the basis of such incentives -- but off the shelf such mechanisms are often central points of failure.} 
\end{itemize}

%% file: s032gain.tex
\label{sec:gain}
In this section we discuss a number of perceived intrinsic architectural advantages to decentralized designs that make them appealing compared to their centralized counterparts.

\subsubsection{Flexible Trust Models}
An intrinsic advantage of decentralized architectures relates to the existence of multiple independent authorities. These create a distributed trusted computing base that ensures that a subset of rogue nodes, at least up to a certain threshold, cannot compromise the overall security properties of the whole system.
 
\vspace{2mm}
\para{Distributed Trust}. Decentralized systems leverage multiple independent authorities into a security assumption: for example, all forms of threshold cryptography~\cite{Shamir79} assure that if some fraction of participants are honest, some security property can be guaranteed. This principle can also be applied to secure multi-party computation, distributed key generation, public randomness and threshold-based decryption, and signing. 
One such privacy system is Vanish~\cite{GeambasuKLL09} that guarantees deletion after a pre-set expiry date. It illustrates how a multi-authority system implements properties otherwise impossible, or implausible, to when implemented by a single entity. However, the system was in practice defeated by a Sybil attack that the security properties of its DHT did not take into account~\cite{WolchokHHFHRWW10}. 
Reliance on multiple authorities to regain a degree of privacy has also been proposed for commercial cloud storage in case some providers are dishonest~\cite{StefanovS13}.

\vspace{2mm}
\para{No Natural Central Authority}. In some settings there exists no central authority and thus a decentralized architecture is a natural choice. This setting has been traditionally studied in the contexts of decentralized access control, as in TAOS~\cite{WobberABL93} and SDSI~\cite{Ellison96}, and `trust management', such as Keynote~\cite{BlazeFK98}. In such systems a set of distributed principals make claims about users and each other, and those claims need to be assembled and used to resolve access control decisions. Bauer et al.~\cite{BauerGR05} show that the task of resolving access control decisions in a decentralized setting is faster than doing so centrally. 

\vspace{2mm}
\para{Leveraging Existing Trust Networks}. In some cases a decentralized infrastructure embeds or expresses a pre-existing set of trust relationships that a system may reuse to support security properties.
Systems may use the underlying social trust structure to build overlay privacy-friendly social network services, as surveyed by Paul et al.~\cite{PaulFS14}. As an example, the Frientegrity system~\cite{FeldmanBFF12} provides a social network platform using untrusted providers seeing only encrypted data, where users can exchange information with `friends' protected by cryptographic access control. This use of encryption to defend against the providers themselves is not the case for systems like Diaspora~\cite{bielenberg2012growth}, an open-source project that takes a different approach: users connect to a provider they trust -- that gains full visibility of their activity -- and delegate the access control on the content they share with their social circles to that provider.

\subsubsection{Distributed Allocation of Resources Assists with Ease of Deployment}

A central premise of P2P networks is that nodes contribute spare resources, and doing away with a central authority that is forced to bear the full costs (such as Google's server costs). This reduces costs and helps ease deployment by spreading these demands amongst multiple parties. Costs are lowered as spare capacity in the existing infrastructure is used, e.g., underutilized resources given by users such as the early SETI$@$home project~\cite{anderson2002seti} and the use of users' storage in Freenet~\cite{ClarkeSWH00}. 

In terms of availability, decentralized architectures exhibit fewer correlated failures by virtue of being distributed. As an example the Cachet system~\cite{NilizadehJMBK12} uses a pool of untrusted peers as a storage back end of a decentralized Online Social Network. 

\subsubsection{Resilience Against Formidable Adversaries}
\para{Location Diversity}. Decentralization provides properties that are inherently difficult to centralize, such as the network location diversity needed for Tor bridges~\cite{dingledine2006design} to bypass censorship both on the network and legal levels. A number of designs take advantage of this, like Publius~\cite{WaldmanRC00}, in order to resist censorship, although censorship resistance itself is a separate field with many centralized, as well as decentralized, solutions. 

\vspace{2mm}
\para{Survivability}. Decentralized architectures can be designed to survive \textit{catastrophic} attempts to take them down or inflict crippling damage, in a way that centralized systems cannot resist~\cite{WylieBSGKK00}. This property has been used to build highly robust botnets using a peer-to-peer architecture~\cite{RossowAWSPDB13}. Although these bot-nets are decentralized on the technical level, they of course maintain central but covert command and control (C\&C). Those botnets have demonstrably been harder to take down using conventional techniques, but are also vulnerable to new threats that result from their decentralization, such as poisoning and enumeration of nodes. A further discussion of wider `Darknet' survivability is provided by Zhou et al.~\cite{li2013finding}.

\vspace{2mm}
\para{Separation of Development from Operations}. Decentralized architectures clearly separate the authorities that provide public code -- and that have no access to operational data and secrets -- and those that run the code. Users and nodes, deploying software, can audit any such open source code for integrity, and chose whether to deploy it. The core development team maintains the code, that is publicly visible and auditable, but upgrading is up to independent relay operators. This model is followed by both Tor and Bitcoin. As a result, attempts to coerce the Tor development team can only have an indirect and possibly highly visible effect -- rendering such attempts less effective. Similarly in Ethereum, the exploitation of a vulnerability in the DAO smart-contract, led to the core developers proposing a ``hard fork'', and this fork was voluntarily adopted by the majority of the Ethereum mining node operators. 

\vspace{2mm}
\para{Publicly Verifiable Integrity.} Due to the availability of multiple independent authorities, decentralized systems can implement accountability mechanisms to publicly verify integrity. Adversaries are disincentivised to compromise nodes, by ensuring attacks have an observable effect so that cheating can ideally be discovered before it has a negative effect. Verifiable logs can be used to help enable privacy as ensuring that actions are transparent enables users to know what happened with their data, as when Pulls et al.~\cite{PullsPW13} use decentralization to support transparent audits of personal data accesses. Auditability is also a key feature of secure electronic election systems such as the Helios system~\cite{Adida08}. Such systems rely on the existence of multiple authorities in a number of ways in e-voting: threshold cryptography is used for parameter and ballot generation, with privacy enforced via threshold decryption.

\vspace{2mm}
\para{Insights}.
\begin{itemize} 
\item \textit{Real-world relationships of trust and authority are personal, complex and localized, and rarely hierarchical or all-or-nothing. Decentralized systems offer flexible trust models that can leverage those relationships to support security and privacy properties.} 
\item \textit{When it comes to high-availability and survivability against powerful adversaries -- particularly with legal authority -- decentralized designs are not just best, but sometimes the only available option. Designs that allow operations to continue despite some authorities being adversarial or not available, are necessary to support these properties.}
\item \textit{Decentralization's fundamental advantage in terms of security stems from an attacker having to compromise a set of independent authorities in order to disrupt or weaken the security properties of a system. Decentralized systems that do not offer this property may be more fragile than centralized equivalents.}
\item \textit{Decentralized designs decouple development from operations and have a multistakeholder governance model, where node operators influence the entire system based on the software configuration they choose to deploy.}
\item \textit{Decentralized systems can leverage public accountability to detect and exclude compromised or misbehaving authorities. Such accountability architectures may be used instead of more complex or expensive prevention techniques, but need to ensure that auditing will be effective and eventually acted upon.} 
\item \textit{Leveraging spare resources of nodes allows decentralized system to scale, and ease deployment. However, this by itself opens the door to high-churn and cannot be a substitute for robust incentives to participate as the system scales or nodes are asked to take on real costs.}
\end{itemize}

%% file: s033privacy.tex
\label{sec:privacy}

In this section we survey the privacy properties obtained through mechanisms that are inherent to decentralized architectures. We limit ourselves to the analysis of technical properties that may be obtained in decentralized systems. We acknowledge that decentralized systems may offer both greater user privacy and autonomous control of the infrastructure. As such they are a possible technological solution to the legally-binding, but often technologically unenforced, demands from data protection laws~\cite{Schaar2010,EDPS2010}, that often are addressed involving a central authority, the data controller~\cite{diaz2013hero}. How decentralized systems relate to the law and business models is out of the scope of this paper.

\vspace{2mm}
\para{Confidentiality from Third Parties}. Some designs employ a decentralized architecture on the grounds that the lack of centralized components, which have full access to user data and can surveil their actions, would be beneficial to confidentiality and unobservability. Such systems may use threshold encryption~\cite{Shamir79} in order to trade off information confidentiality and information availability, such as the PASIS~\cite{WylieBSGKK00} architecture. This scheme splits the data in $n$ ``shares'' and distributes it among peers in such a way that recovering $m$ shares allows one to recover the data, but having less pieces provides no information. Similar solutions are provided by POTSHARDS~\cite{StorerGMV07} or Plutus~\cite{KallahallaRSWF03}.

\vspace{2mm}
\para{Confidentiality from Peers}. In P2P architectures, nodes must interact with other nodes, but they want their communications or actions to remain confidential. For example, nodes need to perform a joint computation, but do not trust each other nor a third party with their data. In this case, decentralization enables them to exchange encrypted data and obtain the sought after result without relying on any particular entity to preserve their privacy. The P4P framework~\cite{DuanYCZ10} is such a system, in which further zero-knowledge proofs are integrated to protect computations against malicious users. More recent, blockchain-backed systems, such as Enigma~\cite{ZyskindNP15} rely more heavily on transparency to achieve this goal. In terms of message-passing, systems that pass end-to-end encrypted messages across untrusted federated servers achieve peer confidentiality. 

\vspace{2mm}
\para{Anonymity.} 
\label{sec:unlinkability}
Due to the distribution of resources in decentralized networks, it is expensive for one entity to observe all actions in the network and track all activities from a user. Many ~\cite{freedman2002tarzan,NambiarW06,McLachlanTHK09,MittalCB12,HerrmannG11}, leverage this approach to provide anonymous communication, although the precise properties provided in terms of anonymity differ.
Some decentralized systems fail to provide full anonymity but instead provide pseudonymity which is weaker ~\cite{PfitzmannH05}, e.g. it allows multiple anonymous actions to be linked, providing weaker privacy, but enabling functionality such as detecting returning users and reducing the complexity of the system. For example, in Bitcoin every transaction is linked to a pseudonym and stored in the blockchain. This allows to trace money flows and avoid double-spending; but on the downside if a pseudonym is ever deanonymized (e.g.~\cite{BiryukovKP14}), all actions from the person would be revealed. A number of decentralized systems, ranging from mix-nets~\cite{Chaum81,danezis2003mixminion}, to DC-nets~\cite{Chaum88}, to Tor~\cite{DingledineMS04}, provide some degree of anonymity.

\vspace{2mm}
\para{Deniability}. Deniability enables a subject to safely and believably deny having originated an action, so as to shield her from responsibility associated to performing such action. The fact that actions cannot be linked back to a user (i.e. ``unlinkability''~\cite{PfitzmannH05}), equips users with freedom to perform actions without fear of retaliation. For instance, in Freenet~\cite{ClarkeSWH00} requests are hard to link to their originator, thus users can freely search for information without revealing their preferences. 

Plausible deniability is crucial in facilitating anonymous and censorship-resistant publishing, and may be implemented using cryptographic techniques allowing of `repudiation'. This was the motivation behind the original Eternity service~\cite{Anderson96} and well-known designs such as Publius~\cite{WaldmanRC00}or Tangler~\cite{WaldmanM01}.

\vspace{2mm}
\para{Covertness}. Some systems protect even the act of participation of nodes in the decentralized network from outside observers (``unobservability''\cite{PfitzmannH05} if the items of interest is the existence of users). In addition to more well-known work like Tor pluggable transports~\cite{pluggable-transports}, the Membership Concealing Overlay Network (MCON)~\cite{VassermanJTHK09} leverages this to provide strong forms of covertness. All nodes in MCON only have links with trusted friends, and a complex overlay network is jointly created that allows all nodes to communicate indirectly with all nodes. As any node only connects to other locally trusted peers, the system defends against attempts to enumerate all users by malicious nodes.

\vspace{2mm}
\para{Insights}.
\begin{itemize}
\item \textit{The key bet of decentralized systems in terms of privacy is that a local adversary may not observe all communications, data, or actions. However, global adversaries are increasingly realistic. Thus decentralized systems that rely solely on dispersion of information to provide confidentiality are fragile.}
\item \textit{Decentralization can harm privacy: Distributing trust and resource contribution to multiple authorities may provide adversarial nodes with extended visibility of user data and network traffic. Thus, naive decentralization designs may in fact create more, not fewer, attack points to breach privacy.}
\item \textit{Decentralization alone cannot balance the needs for privacy, integrity and availability. It is only combined with the use of advanced cryptography that decentralized architectures obtain those properties. In particular, the reliance on others to perform actions, may naturally expose personal information to other nodes without the use of cryptography. However, naive encryption alone may not be sufficient to support the integrity of operations that are more complex than end-to-end messaging.}
\item \textit{Decentralized networks can provide privacy properties like anonymity and even covertness. Yet, most real-world decentralized systems do not use the advanced cryptography and traffic analysis resistance necessary for that purpose as it increases design, implementation, operations and coordination costs.}
\end{itemize}

%% file: s034lose.tex
\label{sec:lose}

Sadly, there is no free lunch in decentralization. While decentralizing has many advantages, there is no guarantee that the properties and features of centralized systems are maintained in the process. This section summarizes problems emerging when decentralizing designs. A further critique of decentralized systems, focusing on personal data, is provided by Narayanan~\cite{narayanan2012critical}.

\subsubsection{Increased Attack Surface} Decentralizing systems across different nodes inherently augments the number of points (attack vectors) that an adversary could use to launch an attack or to observe the users' traffic.

\vspace{2mm}
\para{Internal Adversaries}. In centralized systems, system components can be monitored and evaluated by a trusted entity to detect malicious insiders. In a decentralized system it is easier to insert a malicious node undetected. A number of such attacks have been documented against decentralized systems: the predecessor attack~\cite{WrightALS02,WrightALS04} uncovers communication partners in many anonymous communication schemes~\cite{Chaum88,DingledineMS04,ReiterR98,SyversonGR97}, or the Sybil attack which can be used to bias reputation scores~\cite{Douceur02} or corrupt the information exchanged in collaborative decentralized systems~\cite{JanakiramanWZ03}. Furthermore, when messages are relayed through other nodes, e.g., to gain anonymity, their content is exposed to potential adversaries, as in Crowds~\cite{ReiterR98} for Web transactions or in Yacy~\cite{yacy} for searching information.

\vspace{2mm}
\para{Traffic Analysis}. Decentralization inherently implies that information will traverse a network. Even in the presence of encryption, metadata is available to external adversaries. For instance, in anonymous communications networks it has been repeatedly shown that both passive local~\cite{MittalB08} or (partially) global~\cite{JohnsonWJSS13,MurdochZ07}, as well as active adversaries~\cite{WangCJ05}, can reduce or break anonymity by looking at traffic patterns.

\vspace{2mm}
\para{Inconsistent Views}. Decentralization typically implies that nodes have a partial, thus non-consistent, view of the network which can have an impact on integrity. These non-consistent views allow adversaries to ``cheat'' without being detected. For instance, in Bitcoin adversaries can perform double spending by forcing non-consistency through fast operations~\cite{KarameAC12}, or eclipse attacks~\cite{HeilmanKZG15} in which the adversary gains control over all connections of a target node thus isolating her from the rest of the network.  
Furthermore, the lack of global information results in users not necessarily making the optimal choices with respect to optimizing their privacy, as studied both in the context of anonymous communications~\cite{DingledineM06} and location privacy~\cite{FreudigerMHP09}.

\subsubsection{Cumbersome Management}
An obvious problem of decentralization is that no entity has a global vision of the system, and there is no central authority to direct nodes in making optimal decisions with regard to software updates, routing, or solving consensus.
This makes the availability of a decentralized network more difficult to maintain, a factor significant enough to contribute in the failure of a system, as pointed out by the Mojo Nation developers~\cite{wilcox2002experiences}.
It is very common that nodes in a decentralized system have hugely varying capabilities (bandwidth, computation power, etc.)~\cite{FeldottoSG14,WachsOG14}, making super-nodes attractive targets~\cite{MitraPGG07}. Finally, decentralized systems need to overcome the shortcomings of underlying technologies (such as NAT~\cite{LiuP09}), that favor the client-server paradigm over peer-to-peer networking.

\vspace{2mm}
\para{Defense Difficulties}. The lack of central management hinders the establishment of effective protection mechanisms. For instance, the non-consistent view of the network not only enables attacks, but also hampers the use of collaborative approaches to detect incorrect information~\cite{KapadiaT08}. Similarly, it becomes extremely difficult to prevent Sybil attacks, and defenses must either leverage local information, for example defenses based on social networks~\cite{DanezisM09,YuGKX10}, or collaborative approaches that combine information from several nodes~\cite{ParnoPG05}. 

\vspace{2mm}
\para{Routing Difficulties}. A straightforward consequence of the lack of centralized control is an increased complexity in routing. Nodes do not have an overview of the network and its capabilities~\cite{SuselbeckSKB11} and consequently cannot globally optimize routing decisions~\cite{ZageN07}, falling back to inefficient flooding or gossiping methods in mesh topologies. This is made harder by highly diverse nodes~\cite{FeldottoSG14}, the existence of churn~\cite{ArtigasL09} and the reliance on possibly malicious nodes~\cite{WangMB10}. Solutions to these problems include using complex routing algorithms to enable secure and private discovery of nodes~\cite{McLachlanTHK09,MittalWB13,MittalB09}, or avoiding the use of a centralized directory via next-generation DHTs.
The lack of centralized routing information in decentralized topologies also impacts performance as it hinders the selection of optimal routes or load balancing. We find two approaches to alleviate this problem: using local estimations to improve performance~\cite{AlSabahBG12,AlSabahBGGMSV11,TangG10}, or providing means for users to make better decisions about routing individually~\cite{SnaderB08}. The latter is known to be prone to attacks~\cite{HerrmannG11,MurdochW08}.

\subsubsection{Lack of Reputation}
Decentralization is also an obstacle to the implementation of accountability and reputation mechanisms. 
The negative effect is amplified when privacy and anonymity mechanisms are in place, as it becomes even more difficult to identify misbehaving nodes such as Sybils~\cite{HoffmanZN09}. An effect of this lack of reputation is that nodes have no incentive to behave correctly and can misbehave to obtain advantages within the system (e.g., better performance). This problem has been identified in many settings such as P2P file sharing~\cite{ZhaiCCZLSQTC09}, multicast communication~\cite{YuGS12}, or reputation~\cite{HoffmanZN09}. In particular, the presence of churn, which make nodes short-lived and difficult to track over time, makes the establishment of reputation to guarantee veracity a very challenging problem~\cite{RayaMFH08}, even more if privacy has to be preserved~\cite{SchiffnerPT11}.

\vspace{2mm}
\para{Poor Incentives}.\label{loose-incentives} Without reputation, reciprocity and retaliation it is hard to establish incentive schemes for nodes to not be selfish, in particular in a privacy preserving manner. A solution to this problem is increasing transparency of actions, e.g. by having witnesses to report on malicious nodes in a privacy-preserving manner~\cite{ZhuSJ06b}. However, the most popular approach is the use of (anonymous) payments that incentivize good and collaborative behavior that benefits all users in the network~\cite{BelenkiyCEJKLR07,ChenSC09,KumaresanB14}. In contrast, one example of negative reinforcement is the tit-for-tat strategy to encourage users to share blocks to incentivize sharing, as in BitTorrent.

\vspace{2mm}
\para{Insights}.
\begin{itemize}
    \item \textit{Decentralized designs may prevent conventional attacks but also introduce new ones. Unless they are carefully designed, they may expose personal information to more, rather than fewer parties; and the need to perform joint computation across many authorities introduces threats to integrity.} 
    \item \textit{Decentralized systems are particularly susceptible to traffic analysis, compared with centralized designs, since their distributed operations are mediated through networks and adversarial nodes that may use meta-data to compromise privacy.}
    \item \textit{Decentralized systems by nature require complex management of routing, naming and consistent state -- due to the lack of a central coordinator. Conventional defences against network attacks, like denial of service, require centralization and cannot be straightforwardly applied.}
    \item \textit{Sybil attacks are the great unsolved problem of decentralized systems that allow open and dynamic participation. Solutions based on social networks rely on fragile social assumptions; admission control through identification or payment re-introduce centralization. Proof-of-work defences increase the cost of participation.}
\end{itemize}

%% file: s035assumptions.tex
\label{sec:assumptions}

Even when systems claim to be decentralized, usually there are ``hidden'' centralized assumptions and parts of the design that need to be centralized to operate correctly. These are often implicit.

\subsubsection{Centralization of Network Information \& Computations}
\label{sec:centralizeddirs}
In any decentralized system routing packets across the network is a challenge for both operational and privacy reasons. Typically routing can be divided in two main task. The first task is how to find candidate nodes to relay traffic, and second task is how to select among these nodes. While as detailed in Sect.~\ref{topology}, there are many decentralized algorithms to choose the route, actually finding candidate nodes is difficult, as highlighted in Sect.~\ref{sec:lose}. 
 
\vspace{2mm}
\para{Centralized Directories.} A common solution for the first problem is to assume that there exists a centralized directory that knows all network members. The most prominent example is the Domain Name System (DNS) that resolves easy-to-remember domain names to associated IP addresses in order to allow finding hosts in the largest known decentralized system: the Internet. Though distributed, this centralized service has serious security implications, e.g. for privacy~\cite{MonroseK10} or availability~\cite{VerkampG12}, and thus several alternatives are being proposed~\cite{WachsSG14} and deployed~\cite{DotBit}. Another example are Tor Directory authorities~\cite{DingledineMS04} that provide Tor clients with the full list of onion routers. These directories solve the discovery problem but have become a bottleneck for the scalability of the system~\cite{McLachlanTHK09}. 
How to decentralize these authorities in an efficient, privacy-preserving manner is an active area of research. Solutions are based on having multiple copies of the publicly verifiable directory kept consistent via consensus protocol and distributed via gossiping, although it risks covertness; or to use friend-of-a-friend discovery and routing~\cite{MittalOTBG11,McLachlanTHK09}. 

\vspace{2mm}
\para{Path Selection}. Once routing alternatives are known the question remains: Which route to choose? Thus typically, a centralized server is considered that can ``rank'' routing options to allow for path optimization with respect to adversaries~\cite{AkhoondiYM12,BackesKMM14,EdmanS09,JuenJDBC15}, performance~\cite{SherrBL09,SnaderB08,WacekTBS13}, or with respect to users' reputation~\cite{WangLBH13}. Such a centralized ranking approach has been shown to be vulnerable to attacks~\cite{BauerMGKS07,BiryukovPW13}. Typically DHTs are the possible solution, although only a few have the necessary security and privacy properties for use in decentralized systems~\cite{danezis2005sybil}. 

\vspace{2mm}
\para{Distributed Computations}.A number of decentralized systems are designed with the assumption that there is a central entity that performs computations on the data collected by the nodes in the system. Paradigmatic examples of this behavior are decentralized sensor networks~\cite{ChanP08,EschenauerG02,ZhuSJ06} where the challenge is to send decentralized measurements to a ``master'' node, but there exist other applications such as distributed network monitoring for intrusion detection~\cite{RajabMT05}, anonymous surveys~\cite{HohenbergerMPS14}, or private statistics~\cite{ElahiDG14} in which, even though nodes perform decentralized computations, interaction with a central authority is needed to produce the final result.

\subsubsection{Trust Establishment} 
\label{sec:assumptions:trust}
A challenge when decentralizing networks is to ensure that nodes can be trusted to perform the actions they are assigned or can authenticate themselves as the intended receiver of a message. Often, to avoid dealing with this problem, a common implicit centralized assumption is that a set of trusted servers is assumed to exist, such as in Dissent~\cite{wolinsky2012dissent} or the Directory Authorities in Tor. 

Decentralized trust establishment is still an open problem, though some of the excitement around mining in Bitcoin is precisely due to their attempt to avoid this problem and so build a `trustless' decentralized system.

\vspace{2mm}
\para{Authentication}. In general certificate infrastructures are not decentralized, e.g., PKI. Therefore, some decentralized systems rely on centralized certification authorities to authenticate nodes that can be used for secure routing~\cite{CastroDGRW02,StoicaMKKB01}, user authentication~\cite{BucheggerSVD09}, or to enrol users in the system in the context of anonymous credentials~\cite{CamenischHKLM06,CamenischL01,BelenkiyCCKLS09}, a privacy-preserving alternative for authentication without requiring user identification.
Such centralized authorities are simpler for deployability or usability, but become a single point of failure as pointed out by Lesueur et al. in~\cite{LesueurMT09}. They also introduce an imbalance of power unnatural for decentralized environments since they allow a single entity to revoke peers' authentication credentials. Many decentralized designs do not address authentication (e.g.~\cite{NilizadehJMBK12,SharmaD12}, see~\cite{PaulFS14} for more details), although work from TAOS~\cite{wobber1994authentication} and SDSI~\cite{rivest1996sdsi} onwards has been working in this direction~\cite{macaroons}. Authentication is useful to prevent Sybil attacks, and work on decentralized and privacy-preserving authentication via threshold cryptography is one promising solution ~\cite{maheswaran2013crypto}, as is the use of zero-knowledge systems for anonymous credentials~\cite{BelenkiyCCKLS09}.

\vspace{2mm}
\para{Authorization}.  Assuming the existence of a centralized entity is also common when it comes to storing and enforcing authorization policies, as highlighted by numerous efforts to decentralize policy management and enforcement from SDSI~\cite{rivest1996sdsi} to more recent systems ~\cite{Lesniewski-LaasFSMK07,LiWM03,WinslettZB05}. OAuth was designed to be federated in terms of authorization, but in practice only a few large providers use this standard~\cite{Seong10}. So if an adversary compromises a user's single authentication method such as a password, it can compromise them across multiple decentralized systems. Work descending from SDSI~\cite{rivest1996sdsi} to limited-time authorization via pseudonyms and blind signatures present one way forward to decentralize authorization~\cite{maheswaran2013crypto}. 

\vspace{2mm}
\para{Abuse Prevention}. As mentioned in Sect.~\ref{sec:lose} accountability is a challenge in decentralized systems. Hence, existing abuse-prevention schemes end up relying on centralized parties, often determining global reputation scores. Solutions based on blacklistable credentials (anonymous credentials for which authorization can be selectively revoked) use a centralized authority for enrollment~\cite{TsangAKS07,TsangAKS08}, or to store blacklists~\cite{JohnsonKTS07,TsangKCS11}. Similarly, identity escrow~\cite{BiskupF00} or revocable anonymous communication solutions~\cite{ClaessensDGDPV03}, that allow for re-identification of misbehaving users require a centralized party that stores those identities. In practice, spam prevention in federated email systems also uses centralized lists of known spammers. Typically, these are built from pre-existing trusted social networks, and only recently have reputation systems such as AnonRep (based on homomorphic encryption and verified shuffles) allowed reputation to be done in a privacy-preserving and decentralized manner~\cite{zhai2016anonrep}. 

\vspace{2mm}
\para{Payment Systems}. In many applications of decentralized services it could be desirable to count on a payment system to reward peers for their contributions. While many alternatives have been presented in the literature specifically aimed at peer to peer systems, e.g.~\cite{BelenkiyCEJKLR07,CamenischLM07,YangG03}, they inherently rely on a centralized authority that opens accounts (the bank) and sometimes even on other authorities that can act as ``arbiters'' in case of dispute~\cite{BelenkiyCEJKLR07}, or on authorities that record transactions to help taxation on the operations run in the system, even if the transactions are anonymized~\cite{Taler}. Decentralized crypto-currencies can help ameliorate this problem. 

\vspace{2mm}
\para{Trusted Developer Community}. All decentralized systems work by virtue of having the nodes communicate via the same protocol. Thus, the actual software can be a centralized point of failure if the protocol is flawed. If the protocol is standardized or otherwise uniformly specified, the implementation of the protocol itself may be a failure. Furthermore, the developers themselves could be compromised. his danger is augmented by the software monoculture prevalent in deployed systems, that results in a bug in a popular platform capable of compromising a large set of authorities. One solution is to apply the technique of forcing public transparency and auditing of the integrity of the development process. Open-source development, done in public repositories, is increasingly required. Integrity is ensured via deterministic builds~\cite{reproducible} so that everybody can verify the genuine binary, and the authority to run new versions of the software remains in the hands of the operators. This approach is already followed by Tor and increasingly by Bitcoin, where the choice to deploy particular open-source code is up to miners.  

\vspace{2mm}
\para{Insights}.
\begin{itemize}
    \item \textit{Many decentralized systems implicitly rely on centralized components to hold network information for efficient routing or for establishing trust and defending against Sybil attacks.}
    \item \textit{Essential user-facing infrastructure, from authentication to authorization is centralized even in decentralized systems. Developing alternatives seems to be an open problem, with no clear established design. For payments, Bitcoin has recently provided a decentralized solution, but it suffers from a number of scalability, privacy, and financial volatility problems.}
    \item \textit{The developer community of a system is usually an implicit centralized authority, making social attacks on the developer community itself one of the largest dangers to any decentralized system.}    
\end{itemize}

%% file: s036table.tex

\begin{sidewaystable*} 
\centering
\captionsetup{justification=centering}

\protect\caption{\label{table:systemsTable}Selected decentralized privacy systems evaluated on how they achieve decentralization, the privacy properties they provide, and implicit centralized assumptions.} 

\begin{center}
    \begin{tabular}{l l l  l  l  l  l  l  l  l  l  l  l l}
    &
    System &
    Infrastructure &
    Network Topology &
    Authority &
    \rot{\scriptsize{3rd party-Confidentiality}} &
    \rot{\scriptsize{Peer Confidentiality}} &
    \rot{\scriptsize{Anonymity}} &
    \rot{\scriptsize{Deniability}} &
    \rot{\scriptsize{Unobservability}} &
    \rot{\scriptsize{Centralized Directories}} &
    \rot{\scriptsize{Central Trust Establishment}} \\
    \hline
    \categrow{User Anonymity}
    &\textbf{Tor}$\ddagger$~\cite{DingledineMS04} & Hybrid & Stratified & P2P & \cmark & \cmark & \cmark & \xmark & \xmark & \cmark & \cmark \\
    &\textbf{Mixnets}$\ddagger$~\cite{Chaum81,danezis2003mixminion} & User-independent & Super-Node & P2P & \cmark & \cmark & \cmark & \xmark & \xmark & \cmark & \cmark & \\ 
   & \textbf{I2P}$\ddagger$~\cite{I2P} & User-based & DHT & P2P &  \cmark & \cmark & \cmark & \xmark & \xmark & \xmark & \xmark \\
    &\textbf{Crowds}$\S$~\cite{ReiterR98} & User-based & Mesh & P2P &  \cmark & \xmark  & \cmark & \cmark & \xmark & \xmark & \xmark \\
    &\textbf{MCON}$\S$~\cite{VassermanJTHK09} & User-Based & Mesh & Social &  \cmark & \cmark & \cmark & \cmark & \cmark & \xmark & \cmark \\
    
    \categrow{File Sharing/Censorship Resistance}
    & \textbf{BitTorrent}$\ddagger$~\cite{bittorrent} & User-based  & Super-Node & P2P &  \xmark & \xmark & \xmark & \xmark & \xmark & \xmark & \xmark \\
    & \textbf{Freenet}$\ddagger$~\cite{ClarkeSWH00} & User-based & DHT & P2P  & \cmark & \cmark & \cmark & \cmark & \cmark & \xmark & \xmark \\
    & \textbf{Gnutella}$\ddagger$~\cite{Gnutella} & User-based & Super-Node & P2P &  \xmark & \xmark & \xmark & \xmark & \xmark & \xmark & \xmark \\
    & \textbf{Publius}$\dagger$~\cite{WaldmanRC00} & User-independent & Mesh & Federated &  \xmark & \xmark & \cmark & \xmark & \xmark & \cmark & \cmark \\
    & \textbf{Eternity}$\S$~\cite{Anderson96} & User-independent & Super-Node & Federated &  \cmark & \cmark & \xmark & \cmark & \xmark & \cmark & \cmark \\
    & \textbf{Tribbler}$\ddagger$~\cite{PouwelseGWBYIERSS06} & User-based & DHT & Social  &  \cmark & \xmark & \cmark & \xmark & \cmark & \xmark & \xmark \\
    & \textbf{Vanish}$\dagger$~\cite{GeambasuKLL09} & User-based & DHT & P2P &  \cmark & \cmark & \xmark & \xmark & \xmark & \xmark & \xmark \\
    & \textbf{Tangler}$\S$~\cite{WaldmanM01} & User-independent & Super-Node & Federated & \cmark & \cmark & \xmark & \cmark & \xmark & \cmark & \xmark \\
    & \textbf{Tahoe-LAFS}$\ddagger$~\cite{SelimiF14} & User-independent & Stratified & Federated &  \cmark & \cmark & \xmark & \xmark & \xmark & \cmark & \cmark \\

\categrow{Cryptocurrencies}
&     \textbf{Bitcoin}$\ddagger$~\cite{nakamoto2008bitcoin} & User-based & Super-Node & Ad-hoc$\pm$ &  \xmark & \xmark & \xmark & \xmark & \xmark & \xmark & \xmark \\
& \textbf{Zerocash}$\ddagger$~\cite{Ben-SassonCG0MTV14} & User-based & Super-Node & Ad-hoc$\pm$ &  \cmark & \cmark & \cmark & \xmark & \xmark & \xmark & \xmark \\
& \textbf{MojoNation}$\ddagger$~\cite{wilcox2002experiences} & User-based & Mesh & P2P  &  \cmark & \cmark & \xmark & \xmark & \xmark & \xmark & \cmark \\
& \textbf{Ethereum}$\ddagger$~\cite{ethereum} & User-based & Super-Node & Ad-hoc$\pm$ &  \xmark & \xmark & \xmark & \xmark & \xmark & \xmark  & \xmark \\
\categrow{Secure Messaging}
& \textbf{SMTP+PGP}$\ddagger$~\cite{smtp} & User-independent & Stratified & Federated &  \cmark & \cmark & \xmark & \xmark & \xmark & \cmark & \cmark \\
& \textbf{XMPP+OTR}$\ddagger$~\cite{xmpp} & User-independent & Stratified & Federated &  \cmark & \cmark & \xmark & \cmark & \xmark & \cmark & \cmark \\
& \textbf{Briar}$\dagger$~\cite{briar} & User-based & Mesh & Ad-hoc &  \cmark & \xmark & \xmark & \xmark & \cmark & \xmark & \xmark \\
& \textbf{DP5}$\dagger$~\cite{BorisovDG15} & User-independent & Stratified & Federated &  \cmark & \cmark & \cmark & \xmark & \xmark & \cmark & \cmark \\
& \textbf{Riposte}$\S$~\cite{Corrigan-GibbsB15} & User-independent & Stratified & Federated &  \cmark & \cmark & \cmark & \cmark & \cmark & \cmark & \cmark \\
& \textbf{Dissent/Buddies}$\S$~\cite{wolinsky2012dissent} & User-independent & Stratified & Federated &  \cmark & \cmark & \cmark & \xmark & \xmark & \cmark & \cmark \\
& \textbf{Drac}$\S$~\cite{DanezisDTL10} & User-based & Mesh & Social &  \cmark & \cmark & \cmark & \cmark & \xmark & \xmark & \xmark \\
& \textbf{ShadowWalker}$\S$~\cite{MittalB09} & User-based & DHT & P2P &  \cmark & \cmark & \cmark & \xmark & \xmark & \xmark & \xmark \\
\categrow{Social Applications}
& \textbf{Diaspora}$\ddagger$~\cite{diaspora} & User-based & Stratified & Federated &  \xmark & \xmark & \xmark & \xmark & \xmark & \cmark & \cmark \\
& \textbf{X-Vine}$\S$~\cite{MittalCB12} & User-based & DHT & Social  &  \cmark & \cmark & \cmark & \xmark & \xmark & \xmark & \xmark \\
\categrow{Auditable Systems}
& \textbf{CONIKS}$\dagger$~\cite{MelaraBBFF15} & User-independent & Stratified & Federated$\pm$ &  \xmark & \xmark & \cmark & \xmark & \xmark & \cmark & \cmark \\
& \textbf{Enigma}$\dagger$~\cite{ZyskindNP15} & User-based & Super-Node & Federated$\pm$ &  \cmark & \cmark & \xmark & \xmark & \xmark & \xmark & \xmark \\
& \textbf{Certificate Transparency}$\ddagger$~\cite{laurie2014certificate} & User-independent & Stratified & Federated$\pm$ &  \xmark & \xmark & \xmark & \xmark & \xmark & \cmark & \cmark \\
& \textbf{Helios}$\ddagger$~\cite{Adida08} & User Independent & Super-Node & Federated$\pm$ &  \cmark & \cmark & \cmark & \xmark & \xmark & \cmark & \cmark \\
     \end{tabular}
 \end{center}

\cmark = provides property, \xmark = does not provide property; $\S$ = academic proposal, $\dagger$ = prototype implemented, $\ddagger$ = deployed; $\pm$ = publicly auditable

\end{sidewaystable*} 


%% file: s04directions.tex
\label{sec:directions}

\subsection{Address Decentralization's Shortcomings}\label{sec:recover}

To build the next generation of decentralized systems, good will, slogans, and demands are not enough. What is needed is a clear research plan. A number of designs we review consider decentralization as a goal and virtue in itself and do too little to address the inherent challenge of maintaining privacy properties and deployment with high availability. In particular we studied in Section~\ref{sec:lose} a number of those challenges: an increased attack surface, with corrupt insiders; susceptibility to peers violating privacy and vulnerability to traffic analysis, integrity and consistency attacks; expensive and fragile routing; potential degradation in performance; loss of central choke points to enforce security controls; peer diversity and lack of incentives. These are serious and real threats, and not acknowledging them and confronting them head on leads to weak systems that cannot credibly compete with centralized solutions. This is demonstrated by the failure of Ethereum to promptly address the DAO vulnerability~\cite{DAO}. Indeed, decentralization in the style of early BitTorrent simply ends up being an inefficient way to do redundancy and availability without a centralized authority --- and with no credible privacy properties. Likewise, Bitcoin and Ethereum provides this style of decentralization with the addition of integrity but their simplistic accountability designs harms privacy. Therefore, more research is required looking at systems such as Tor and Bitcoin as platforms rather than purely as channels, including understanding their interfaces, performance, quality of service guarantees and the privacy properties as a whole system in order to deliver better privacy properties. 

Availability without centralization is a key promise of decentralized systems, but often fails when the system grows. The most important engineering challenge of those reviewed is that decentralized systems often do not scale and are inefficient in comparison to centralized systems. In practice, in a world with limited resources and investment, inefficient decentralization leads to a failure of decentralization. This problematic dynamic is built into decentralized designs: maintaining high-integrity requires a majority to honestly participate in decisions. Although one could point to Bitcoin as a success, the larger Bitcoin network of miners grows the less it scales, as all miners need to detect and verify new blocks and transactions. Even worse, Ethereum smart contracts are executed on each node in the network. In both Bitcoin and Ethereum, as the number of nodes grows, the system gets slower. Due to this unfortunate design flaw, Bitcoin and Ethereum will face serious issues when scaling without major design changes that accountability as such does not address. We can be assured the current generation of attempts to ``re-decentralize'' the Internet will fail without more research on how to scale efficiently.

Finally, there has to be a deeper acceptance that even honest users and peers in decentralized systems will have to be incentivised to participate and behave cooperatively. This is particularly true when stronger privacy protections are implemented and reputation based on repeated and iterated interactions cannot be leveraged. In those cases standard platforms must be developed to prevent Sybil attacks and establish privacy preserving reputation to curtail abuse; accounting and payment mechanisms need to be devised to ensure that those that do work are rewarded to sustain their operations. Systems that do not provide incentives for participation in the infrastructure will fall foul of the tragedy of the commons and will remain mere proofs of concepts. 

Even with motivated users, human fallibility must be addressed realistically. Decentralization advocates desire of users to return to a `lost golden age' of self-hosting services, as in the `re-decentralize' project~\cite{redecentralize}. However, the popularity of services like Facebook and Gmail shows that most do not have the time or skills to host decentralized nodes unless a powerful incentive exists such as file-sharing. 
Worse, users may not be qualified at protecting their own systems, when even most skilled professional administrators cannot. Building successful decentralized systems that do not betray the security and privacy of their users is hard, and entails much more than tacking a blockchain or P2P network to a pre-existing problem, but also has to take into account platform security and ease of user operations.

\subsection{Develop Design and Evaluation Strategies}
Systems that claim to be decentralized today simply often use the adjective in an informal manner, resulting in decentralized ``snake oil'', as is the case for some blockchain-based start-ups. Unlike formal security definitions, information-theoretic definitions of anonymity, and differential privacy, there are no coherent quantitative metrics to characterize decentralization. Aside from having a common definition of the privacy and security properties, decentralization engineering also requires the development of design strategies that measure both decentralization and its effect on the properties systematized earlier.  More often than not, properties are neglected, rarely mentioned or evaluated, including the impact of decentralization and availability. Section~\ref{sec:how}, for instance, illustrates the variety of options in this design space. 

Beyond the impact of decentralization on availability, a key missing piece is a systematic means for evaluating the privacy and security properties provided by a given decentralization system. As we evidence, decentralization can support privacy in many ways (Section~\ref{sec:privacy}), as well as supporting other properties too (Section~\ref{sec:gain}). We observe that systems are often designed with one particular privacy goal in mind, which is frequently redefined to suit the design, and system designers tend to resort to ad-hoc evaluation. A particular case in which a lack of systematic evaluation has great impact in terms of understanding the protection provided by decentralized system is the case of compound systems (i.e, systems that combine different schemes to try to improve overall protection); or the case where systems are deployed in environments with different characteristics than those assumed in their design. In decentralized systems, it is not granted that the protection of the whole is greater or equal than the sum of the parts. In fact, the inverse may hold: combining different decentralized systems with different assumptions may violate the properties each system guarantees by itself. For example, while a user may assume using BitTorrent over Tor provides anonymity for file-sharing, in fact the reverse holds: Tor provides no anonymity to UDP-based systems like BitTorrent, and users can even be deanonymized by virtue of running BitTorrent~\cite{le2010anonymizing}. In other words, systems to not exist in a vacuum. Their analysis and evaluation needs to account for interactions with their environment or other systems.

A similar trend is observed in terms of measuring the severity of disadvantages introduced by decentralization. Though, as we show in Section~\ref{sec:lose}, many weaknesses arise from decentralizing, few works evaluate their implications, or do so in a design specific way that is difficult to extrapolate to other systems. As a result it is extremely difficult to compare systems and find promising new directions. This slows the development of robust decentralized systems by obscuring good design decisions. For example, in many systems there is a trade-off between privacy and availability. 

Further work is also required to radically simplify the deployment and management of ``real-world'' decentralized applications, either on larger platforms or as stand-alone distributed systems. Deployability and usable application life-cycle support is at the heart of the current centralized cloud-based `dev-ops' revolution, and has made centralized app stores and Web applications as popular as they are. Yet, there are no equivalent tools or technologies to facilitate the deployment, management, and monitoring of decentralized systems, let alone their continuous updates, application life-cycle management, and telemetry. This gap negatively affects developer's productivity and makes the engineering and maintenance of decentralized systems very expensive. Building toolchains that support easy management -- without introducing any central control -- is largely an open research problem. Successful projects such as Tor and Bitcoin have developed best practices and running code in that space such as open-source development and \textit{reproducible builds}~\cite{reproducible} to address security concerns that may be generalized.

\vspace{2mm}
\para{Key Research Questions for Decentralization}.
\begin{itemize}
\item \textit{Are there generalized techniques to provide privacy and integrity properties for decentralized systems without damaging availability?}
\item \textit{Can we develop systematic techniques to evaluate decentralized systems both in isolation and when they are deployed in different environments?}
\item \textit{How can human users be incentivised to work in a decentralized manner?}
\item \textit{How do real-world deployment of decentralization lead to scalability challenges that change the desired properties and defeat decentralization?}
\item \textit{Can we develop a mathematical metric to define degrees of decentralization?} 
\end{itemize}

In the next section we will provide provisional answers to these questions to guide future research. These answers will be based on the observations built in previous sections.

%% file: s05conclusions.tex
\vspace{2mm}
\para{Availability, Privacy, and Integrity}.
Our analysis points to some fundamental trade-off between availability, privacy, and integrity in decentralized systems: A good design for one is an unsafe design pattern for another. Systems use a wide variety of infrastructure, network topology, and authority relation choices (as systematized in Table \ref{table:systemsTable}). Three widely deployed decentralized systems demonstrate a different set of design goals. Bitcoin comes with high-integrity at the cost of a public ledger with little privacy. Tor routers provide high-privacy at the cost of no available or correct collective statistics to ensure the integrity of the entire system. BitTorrent provides high availability in downloading files, but fails to provide privacy to its users against powerful adversaries. 

We believe it is not pre-ordained that there is a trade-off between privacy, availability, and integrity in decentralized systems by virtue of using advanced cryptographic techniques. Unlike Bitcoin, Zerocash\cite{Ben-SassonCG0MTV14} combines both privacy and integrity using zero-knowledge proofs. Likewise, many academic systems, such as Drac\cite{DanezisDTL10}, tackle traffic analysis to defend privacy in a P2P network. Simply put, advanced techniques for providing everything from dummy traffic for anonymity to succinct zero-knowledge proofs are not yet part of the toolbox for many decentralized system engineers.  

\vspace{2mm}
\para{Interdisciplinarity}.
Reviewing the literature reveals that to build good secure privacy-preserving decentralized systems, one needs:
\begin{itemize}
\item Expertise in building \textit{distributed systems}, as decentralized systems are by definition distributed.
\item Knowledge of modern \textit{cryptography}, as complex cryptographic protocols are necessary to achieve simultaneously privacy, integrity and availability.
\item An understanding of \textit{mechanism design, game theory and sociology} to motivate cooperation amongst possibly selfish actors.
\end{itemize}

The focus on social incentive structures is usually left out, and thus most decentralized systems do not gain real-world wide deployment. In general, the involvement of nodes in decentralized systems varies and this is usually mirrored in the power allowed to authorities, as well as in inter-node relationships that reflect social behavior. Some designs assume centralized components, for better availability and performance. Others push for sheer decentralization, in pursuit of resilience to censorship and network outages. Are these design choices often social or political rather than technical? Most designs, though, fall somewhere in the middle and generally impose cryptographic techniques and rely on real-world dynamics in order to defend against adversarial nodes. Certainly, the way decentralization is achieved affects the privacy of the users and thus their behavior. It falls upon decentralized system designers to achieve satisfactory performance and deployability, while taking into account not just the technical but the necessary social structure of the system.  

\vspace{2mm}
\para{Real-world Scalability}. From our study of the literature, we have shown that a number of key functions of decentralized systems often fall-back to centralized models in practice for scalability, even when unnecessary. First, network directories, key management, and naming often remain centralized. Thus, the there is a need to design of collective high-integrity and re-usable infrastructures to support directories, node discovery, and key exchange. These mechanisms need to scale up and remain decentralized, while not being open to corruption or inconsistencies. 

Second, reputation and abuse control often require either centralized entities, or building on pre-existing social networks in user-based infrastructure. Even advanced privacy-preserving techniques, such as anonymous blacklisting, assume that centralized services will issue and bind identities, and e-cash protocols rely on a bank to issue coins and prevent double spending. More work is required in establishing reputation in decentralized systems and preventing abuse without resorting to central points of control.

Third, it is important to make credible assumptions about the platform security and computing environment of end-users or other devices. It is too facile to heavily rely on end-user systems keeping secret keys and data, and ignore that they are often compromised. Achieving perfect end-point security is an ambitious goal in and of itself -- and so needed but beyond the strict remit of building secure decentralized systems. Decentralized architectures that display or limit the effect of compromises, and which may `heal' and recover privacy properties following hacks, should be preferred to those that fail catastrophically or silently under those conditions.

\vspace{2mm}
\para{Defining Decentralization}. In general, decentralized systems are networks. Yet as shown by the difference between network topologies for routing and the relationships of authority, a decentralized network is not simply a single network, but multiple kinds of networks connected on different levels of abstraction. Worse, the overly simplified models of decentralization presented in many papers and research prototypes do not take into account the changes produced by real-world usage into account. As shown by BitTorrent, simple decentralized networks tend to evolve from P2P into super-node systems. In general, as a system scales there is a tendency towards distribution, but not decentralization, in order to maintain efficiency. Using network science, one can show simple models such as random graphs with basic mechanism design such as preferential attachment scale into small-world systems over time, and these systems often simply transform into a federated client-server architecture or a simple centralized distributed system. In order to maintain decentralization as an emergent property, it appears that advanced hybrid and stratified system, e.g. Tor, are necessary to ``unnaturally'' maintain decentralization and the relevant privacy properties. Yet, the Tor network has many centralized technical (complete network information by directory authorities) and social assumptions (control by a core group of developers). The key point of a real measure of decentralization should be to take these more stratified designs into account. An ideal decentralized system would remove all centralized assumptions while maintaining the needed security and privacy properties. 

The ultimate bet of decentralized systems is still open: is being vulnerable to a (possibly random) subset of decentralized authorities better than being vulnerable to a single centralized authority? Decentralization seems to be the result of a breakdown in trust in centralized institutions, but we do not yet understand how to build decentralized social institutions to support decentralized technical systems despite the promises of Bitcoin to produce algorithmic monetary policy, or the promise of Ethereum to support modern civilization with scripts with dubious security properties. Decentralization is a hard problem, but the fact that it is technically amendable to advanced techniques from distributed systems and cryptography should indicate that the social questions at the heart of decentralization are not unsolvable.